# *Visualization and thermodynamic encoding of single-molecule partition functions*

Carlos-Andres Palma[1]*, Jonas Björk[2], Florian Klappenberger[1], Emmanuel Arras[1], Dirk Kühne[1], Sven Stafström[2] & Johannes V. Barth[1]*

1. Physik-Department E20, Technische Universität München, D-85748 Garching, Germany
2. Department of Physics, Chemistry and Biology (IFM), Linköping University, Linköping 58183, Sweden

**Abstract**

Ensemble averaging of molecular states is fundamental for the experimental determination of thermodynamic quantities. A special case occurs for single-molecule investigations under equilibrium conditions, for which free energy, entropy and enthalpy at finite-temperatures are challenging to determine with ensemble-averaging alone. Here, we provide a method to access single-molecule thermodynamics, by confining an individual molecule to a nanoscopic pore of a two-dimensional metal-organic nanomesh, where we directly record finite-temperature time-averaged statistical weights using temperature-controlled scanning tunneling microscopy. The obtained patterns represent a real space equilibrium probability distribution. We associate this distribution with a partition function projection to assess spatially resolved thermodynamic quantities, by means of computational modeling. The presented molecular dynamics based Boltzmann weighting model is able to reproduce experimentally observed molecular states with high accuracy. By an *in-silico* customized energy landscape we demonstrate that distinct probability distributions can be encrypted at different temperatures. Such modulation provides means to encode and decode information into position-temperature space or to realize nanoscopic thermal probes.

**1 Introduction**

Statistical thermodynamics is one of the pillars of the atomistic theory of matter[1-3]. In this context, the partition function plays a central role, bridging the distribution of states in a given system with macroscopic quantities, such as the free energy or specific heats. Because of the astronomically large number of available microstates in a typical many-molecule ensemble, extracting thermodynamic quantities through partition functions is elusive for all but the simplest of molecular systems. Alone for a single molecule, the configurational partition function[2,4], depending on the positional part of the spatial degrees of freedom and the potential energy only, can be expressed as:

$$Z = \sum_{i}^{N} \exp(-\beta U_i)$$

provided that each microstate *i* corresponds to a uniquely defined configuration $x_1,...,z_j$ and that the configurational space *N* is complete.

The advent of nanoscale science offers to drastically reduce the configurational space *N* by molecular constraining. The most striking achievements have been obtained by intricate measurements at non-equilibrium conditions. For instance, by stretching macromolecules with a sharp probe or molecular tweezers[5-9] data have been obtained that can be related to single-molecule thermodynamics[10-12]. In addition, other nanoscale investigations were reported implying ensemble-averaging of many-molecule distributions[13-15]. The commonality of these examples is that they rely on ensemble- or time-averaging a collection of observable macromolecular features that are fundamentally sub-molecular in nature. Therefore, microscopic thermodynamic information, i.e. microstate probabilities with atomic precision, is usually lost[16,17]. However, it is clear that the exploration of the spatial degrees of freedom underlying the configurational partition function can be expressed geometrically for a given system.

From the modeling point of view, molecular dynamics (MD) simulations currently provide a central approach to compute microstate probabilities with atomic precision[18]. With the use of classical force fields and emerging strategies such as network projections[19], MD allows computing equilibrium microstate probabilities with millions of microscopic molecular states and microsecond timescales, thereby offering the possibility of full convergence to the ergodic limit. A drawback in MD modeling is the dependence on force-field validation, which limits its use to well-known systems. Because of this inherent restriction, a common strategy in single-molecule experiments has become to calculate zero-temperature potential energy surfaces through *ab-initio* methods and assume their relevance for finite-temperature experiments[20-23]. Altogether, neither experimental nor simulation techniques have tackled temperature-dependent free energies of single molecules in well-defined equilibrium environments with sub-molecular resolution. The ability to do so provides not only fundamental insights and multiple prospects for single-molecule thermodynamics, but also may be regarded as basis for new types of sensors, computing and encryption protocols in molecular science.

Herein we make use of surface-confined nanoporous metal-organic nanomeshes (MONs) on a weakly corrugated Ag(111) surface to explore the equilibrium thermodynamics of single caged species at equilibrium conditions. The generic recipe for the preparation of 2D MONs is the deposition of multitopic ligands on an atomistically clean planar substrate (e.g. silver, copper) followed by evaporation of a transition metal (e.g. cobalt, iron) and annealing under ultra-high vacuum (UHV) conditions to induce metal-directed assembly of coordination networks[24-29]. MONs provide versatile scaffolds to confine atoms or molecules[30-32], tune interfacial electronic landscapes[30,32] and steer metal growth[24]. They constitute 2D analogues of the broader field of three-dimensional reticular and framework chemistry[31].

The dynamic behaviour of organic molecules at homogenous surfaces has been extensively studied and deep insight gained in their mobility characteristics. By scanning tunneling microscopy (STM) investigations molecular-level characterization of individual diffusing species became possible[33]. Notably the translation or rotational motions of aromatic flat-lying species could be followed in exquisite detail, and evidence appeared that the formation of supramolecules or nano-architectures leads to special mobility scenarios where intermolecular interactions sensitively interfere [22,34,35].

Through adequate assembly protocols we prepared MONs with mainly single molecules captured in the hexagonal pores. The STM image and model in **Fig. 1** depict a porous network structure defining a regular honeycomb superlattice from *para*-سexiphenyl-dicarbonitrile with threefold lateral coordination to Co centers. By caging an additional rod-like single linker, a system is at hand whose dynamics can be followed in detail by temperature-controlled measurements, because the regular shape, nm-extension and aspect ratio of the molecule are favorable for identification in STM data (all images presented in this work exclusively show nanopores in which a single molecule is trapped). Note that the length of the molecule (29.6 Å) is comparable to the size of the network pore (rim-to-rim distance of 58 Å).

Upon positioning a guest species in the regular nanopore of a MON, the 2D thermal motions become restricted to a patch of nanometre-dimensions (specifically, to the ~24 nm$^2$ van der Waals cavity of the network for the system under investigation[27]), and at the same time influenced by the presence of the nanomesh rims that represent unsurmountable walls provided the thermal excitations are reasonably low. As a consequence the number of available microstates of the system reduces dramatically, opening the possibility to perform simulations at the ergodic (predictive) limit. Previous STM studies[26,27] of 2D MONs on Ag(111) surface under UHV conditions consisting of cobalt-coordinated dicarbonitrile oligophenyls[36] revealed caged supramolecular dynamers in propeller-like trimeric configurations[28]. Interestingly, the findings testified reversible switching from a kinetically trapped six-star pattern at 87 K to a thermodynamically stable twelve-star pattern at 145 K. Such pattern modulation suggests that specific pattern coding via energy landscape design and encryption through temperature control is possible. This offers the possibility of encrypting/decrypting information by thermodynamic design alone, i.e. at the ergodic limit. Indeed a thermodynamically encrypted pattern is uniquely defined and more predictable than a kinetic one.

Herein we exploit the molecular confinement as a platform to directly probe time-average patterns (TAPs) expressing a multitude of single-molecule states[37]. Importantly, with the reduced space, *ab-initio* parameterization of molecular force fields is possible, allowing MD sampling and extraction of equilibrium microstate probability distributions with atomic precision. To tackle this issue, we implement a projection of the configurational partition function in real space (which effectively constitutes a projection of the microstate probability distribution[38]) and establish its correlation with the experimental TAPs. We use this approach to quantify ergodicity at the nanoscale, computing free energy differences between experimentally accessible TAPs and simulations amounting to 0.2 cal mol$^{-1}$ K$^{-1}$ or 0.5 kcal mol$^{-1}$ at 250 K and with sub-molecular resolution. This quantification allows by-design (predictive[39]) thermodynamics. We show how a straightforward design of a potential energy landscape can be used to encrypt/decrypt information in (*x,y,T*)-space at the ergodic limit. The findings and concepts provide intriguing analogies to the recently reported holographic encoding based on surface-confined fermionic states designed by molecular manipulation protocols[40].

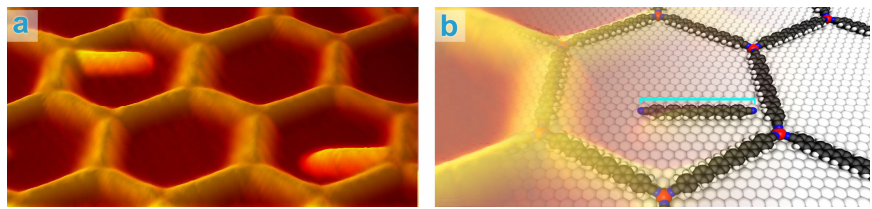

**Fig. 1. Confinement of single molecules in a surface-confined metal-organic nanomesh. a)** Perspective STM image recorded at 8 K showing the static configuration: in two of the honeycomb pores host an individual immobilized sexiphenyl dicarbonitrile molecule has been trapped (emphasized by light color). **b)** Model showing the registry of the metal-organic superlattice with the underlying Ag(111) substrate (Co centers with lateral threefold coordination to CN groups are shown in pink). The caged unit (with N-to-N distance of 29.6 Å indicated in cyan) is oriented along high-symmetry substrate directions and attached to the nanomesh rim by weak noncovalent bonding. The nanopore's outer diameter is 67 Å.

## 2 Equilibrium states and dynamics of caged single molecules

The guest species' thermal motions are frozen at T=8±1 K, under which conditions low-temperature STM data reveal (**Fig. 2a**) their preferred attachment via a CN group to the MON honeycomb's rim close to the vertices (cf. **Fig. 1 b**). The backbone orientation follows low-index <1-10> substrate directions and for symmetry reasons twelve such equivalent configurations coexist. With slightly higher temperatures a 1D guided diffusion[22] sets in, as illustrated by the image sequence in **Fig. 2b-f** obtained at 28 K. During this lateral motion the guest molecules follow the phenyl backbone of a honeycomb segment while keeping their orientation. Intermediate states between the preferred corner positions often imply a fractional imaging (**Fig. 2c,e**) of the guest, due to the reduced residence time of these states being shorter than the minimum measurement period with the employed scanning frequency and area. Accordingly, guests in corner positions are associated with the highest occurrence probability, which similarly appears in a statistical analysis of an ensemble (cf. **Fig. 2h-l**). Upon further increase of the temperature a fundamentally different imaging regime unfolds, reflecting a high mobility of single guests, which is clearly distinct from related effects with caged supramolecules, cf. supplementary information (SI) **Fig. S0**, annexed. This is illustrated by the STM data, recorded at 82±1 K, depicted in **Fig. 2n,o**: now the same characteristic pattern with six-fold symmetry is identified in all occupied pores. From the thermodynamic viewpoint, the caged species now perform both translational and rotational motion in the pore. Thus, rapid intrapore thermal motions including transient occupation of accessible states identified at lower temperatures occurs, with hopping rates largely exceeding that of the fast-scanning direction (typically 4 Hz). Accordingly the imaging process renders an intrapore corrugation pattern with intensity maxima retracing the preferred configurations.

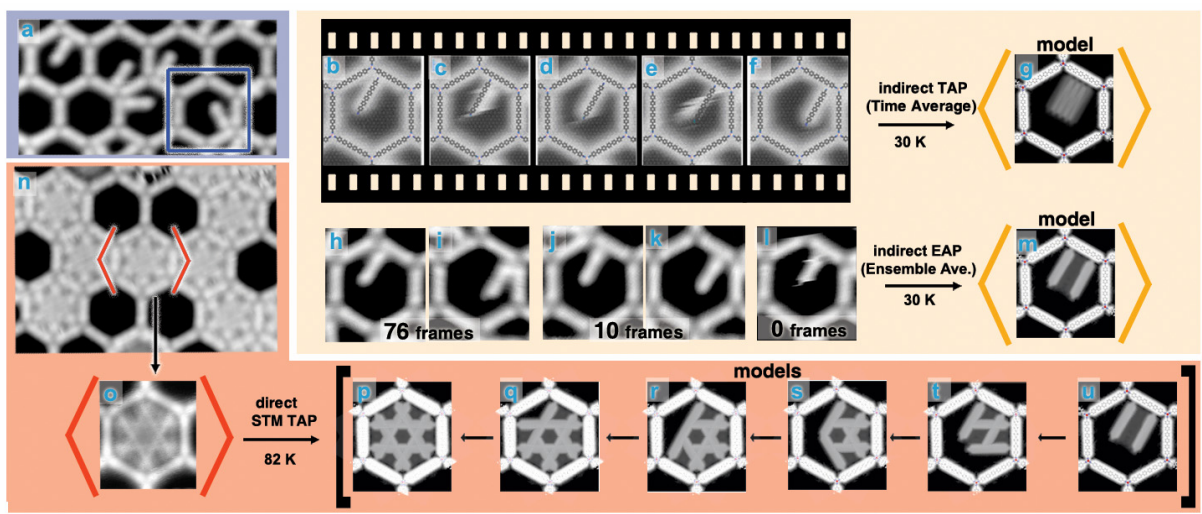

**Fig. 2. STM imaging and thermodynamic methods addressing caged single molecules at variable temperatures** (8 K (blue) 30 K (yellow) and 82 K (orange)). **a)** STM image of static configurations with caged immobilized single guests in the hexagonal pores at $T_{Sample}$ = 8±1 K. **b-f)** STM image sequence following the guided 1-D thermal motion of a guest along a honeycomb rim in a selected pore at $T_{Sample}$ = 30±1 K, associated with a path between two equivalent microstates. Models are superimposed (tunneling parameters $V_B$ = 2.0 V, $I_T$ = 0.1 nA; average time lapse 720 s. **g)** Time-averaged pattern of the previous sequence using molecular models indirectly rendered through Equation 1. **h-k)** Statistical analysis of overview STM data recorded at 30±1 K, identifying different microstates and their occupation frequency. **l)** Transient configuration identified in a close-up measurement, elusive in large-area surveys. **m)** Ensemble-averaged pattern (EAP) of a distribution of 86 molecular models with the microstate weights as shown in **h-k**, indirectly rendered using Equation 1. **n)** High intrapore mobility at $T_{Sample}$ = 82±1 K: rapid diffusion with preferential occupation of microstates identified at lower temperatures lead to a characteristic STM imaging pattern with six-fold symmetry. **o)** High-resolution single-molecule time-average pattern (TAP) in an individual pore at 82±1 K: the main features are reproduced by the overlay of consecutive 30±1 K EAP in m, (**p-u**).

## 3 Direct vs. indirect thermodynamic averaging in real space

Low-temperature STM imaging is often used to reveal microstates, i.e., single-molecule configurations corresponding to local energy minima. At such cryogenic conditions, the statistical occurrence of the different microstates not necessarily represents an equithermal molecular distribution, due to possible kinetic trapping of intermediate microstates[39]. One apparent solution to this problem would be a statistical analysis at a temperature where transitions between the intermediate microstates of interest occur such that equilibration of the molecular ensemble becomes possible. In our system, these transitions set in at T ≈ 30 K as demonstrated by the experimental STM time frames in **Fig. 2b-f**, where a caged molecule diffuses along a honeycomb segment from an initial position designated **G,** to the opposite **G'** (*vide infra*). It is illustrative to represent the situation in graphical form as a TAP, which can be *indirectly* constructed by dividing the pore space in a pixel-wise fashion. Note that *indirect* refers to the mathematical reconstruction of an observable from isolated samples. By averaging the pixel-wise occupation probability *p* we have:

$$p_{pixel} = \frac{1}{M} \sum_i^M s_i \qquad (1)$$

where $M$ is the number of the considered time frames. The parameter $s$ equals 1 if a molecule populates the volume represented by the pixel and 0 otherwise. The example in **Fig. 2g** depicts the so-constructed TAP using the very limited serial data set of **Fig. 2b-f**, whence we find the same occupation probability for the five apparent microstates along the pore segment instead of the proper statistical weights, i.e., because of the relevant time scales, pore state equilibrium properties were not extracted reliably. According to the ergodic theorem, a second strategy consists in analyzing extensive data sets to construct the ensemble-averaged pattern (EAP) at the same temperature by the indirect means of Equation 1, where $M$ now refers to the number of inspected pores. For large numbers, all relevant microstates should be observed with the appropriate statistical weights. When 86 single-molecule occupied pores are measured at 30±1 K and analyzed, we find that 76 are distributed among microstates **G** and **G' (Fig. 2h,i)**, while 10 correspond to intermediate microstates **I** and **I' (Fig. 2j,k)**, i.e., there is an occupation probability of 0.88 for **G,G'**. The rendering of the pertaining EAP is illustrated in **Fig. 2m** and captures essential statistical characteristics of the system. However, a transient microstate **C** that could be identified in close-up measurements (cf. **Fig. 2l**) proved elusive in the overview scans since it is too short-lived. These limitations demonstrate that for addressing single-molecule thermodynamics of molecular species in caging environments indirect TAP or EAP averaging methods may be insufficient or even misleading.

For an increased acquisition frequency of microstate statistics, raising the temperature is an efficient means to boost diffusion rates. Confining the single molecule to a specific nanoscopic area is hereby essential for keeping the accessible microstate space constant. Thus, ergodic microstates sampling within a well-defined environment becomes possible and we can *directly* measure TAPs representing equilibrium properties. For our system a wide temperature range (between 70 and 145 K) exists, where the dynamic behavior of single caged molecules produces quasi-static topographies for the reasons addressed above and all occupied pores show exactly the same TAP. An exemplary situation is depicted in **Fig. 2n**, where seven filled pores were simultaneously monitored at 82±1 K, all exhibiting the same distinct hexagonally symmetric pattern. The high-resolution data (**Fig. 2o**) reveals that its main features can be mimicked by symmetry operations obeying the signature of the nanopore (**Fig. 2p-u**, where the contrast is normalized for clarity) applied to the previous indirect EAP model (**Fig. 2m**). These experimental patterns present a *direct* visualization concerning the projection of the pertaining molecular configurational partition functions and notably reflect both 2D translational, rotational, and part of the vibrational motions of the caged molecules under equilibrium conditions, whence we can interpret them as a *projected* partition function for individual caged molecules in the framework of Boltzmann statistics[1,3,41] (*vide infra*). It is worth mentioning that the classical configurational partition function does not take into account electronically excited states and the treatment in the next sections supposes molecules in their ground electronic configuration. This is a good approximation considering the experimental conditions (T < 150 K).

## 4 Energy landscape and modeling

The analysis of STM results indicates an energetically favorable microstate with the confined molecule near the pore vertices. The high-resolution STM image with the atomic models for the underlying Ag(111) (cf. **Fig. 3a**, silver atoms as dark gray spheres), the Co adatoms (red) and molecules (black, white and blue) shows how this 'ground' microstate **G** aligns along the surface lattice in a <1-10> direction. For a model description of the system by MD simulations at finite-temperature, high-quality force-fields have to be developed. Accordingly, an all-atom MD model of the pore was constructed with customized force-field (FF) parameterization (see SI **Fig. S1-S7**). Initially, the electrostatic potential of the empty pore was parameterized to the density functional theory (DFT) electrostatic potential of the full unit cell (**Fig. 3b**). Subsequently, the ground state geometry was calculated with DFT, reproducing the experimentally observed microstate **G** (**Fig. 3c**). Then, the molecule was translated in perpendicular orientation along the honeycomb segment, while keeping a constant distance of 2.18 Å between the rim atoms and the nitrogen of the guest at an adsorption height of 3.0 Å. For reproducing the energy landscape, the force field's electrostatic parameters and *ad hoc* image charges were fitted (**Fig. 3d**, red line see **Supp. Note 1**) to match the DFT results (**Fig. 3d**, blue line). Note how the DFT and FF landscape sets the scene for the STM observations depicted in **Fig. 2h-k**: the second-most favorable microstate is the 'intermediate' microstate $I_2$, lying 20 meV above the **G** microstate. Intriguingly, **Fig. 3d** also revels how the microstate minima (red arrows) closely follow the surface corrugation rather than the non-covalent interactions with the pore rim. Nevertheless, the energetically favored **G** microstate seems at an ideal position between biphenyl hydrogens as previously reported for the underlying non-covalent interaction scheme between CN groups and phenyl rings[42,43].

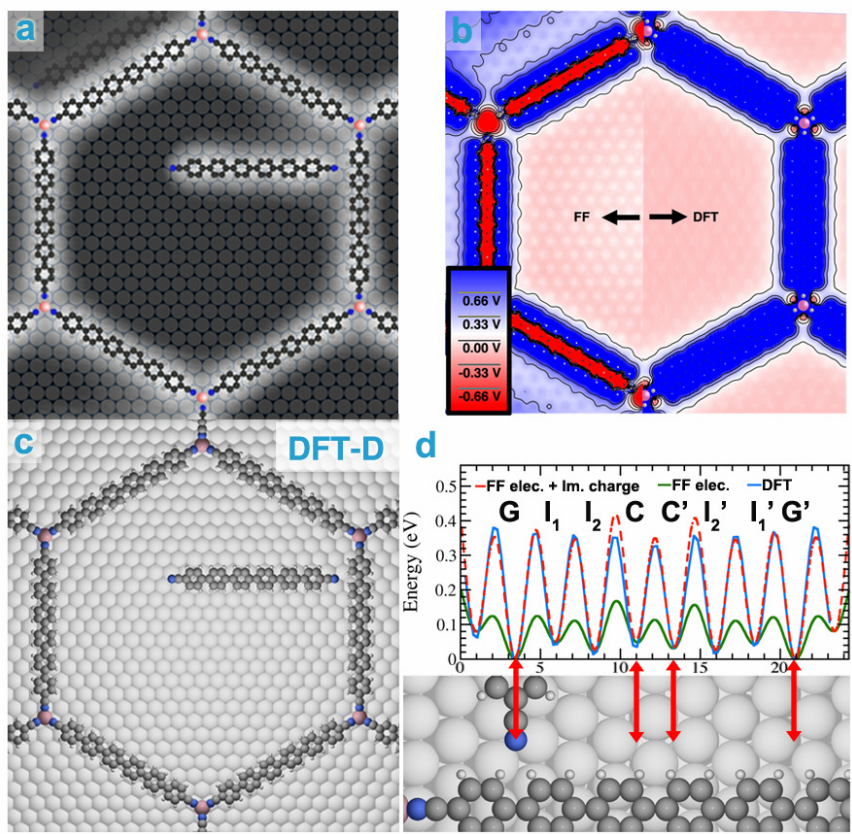

**Fig. 3: Energy landscape of single guest translated along honeycomb rim. a)** STM image at 8 K showing the favored microstate of a caged molecule. **b)** Electrostatic potential slab at a height of 3 Å from the Ag(111) obtained from solving the Poisson equation for a DFT generated density and from force-field point charges (FF). The white (neutral charge) areas inside the pore correspond to the position of the Ag(111) top sites. **c)** The fully relaxed molecule at the G microstate using DFT-D level of theory. **d)** The force-field (red and green lines) and DFT (squares, blue line as a guide) energies of a molecule translated across the path shown in Fig. 2b-f with a distance of 2.18 Å between the terminal N and the pore rim.

## 5 Image free energy and *projected* sum over states for restricted translation

Next, we introduce a method to model and analyze the TAPs in a local, pixel-wise fashion by MD simulation sampling. As a simple test case the guided diffusion along a single decorated pore rim is considered, i.e., we hypothesize a simple 1D dynamic regime. This essentially implies that an equilibrium situation exists corresponding to the experimental EAP at 30±1 K(cf. **Fig. 2m**). Note again, such illustrative 1D diffusion scenario cannot be addressed by an STM TAP (imaging at slightly higher temperatures causes spurious 2D intrapore diffusion pathways). By using **Equation 1** and two MD relaxed frames ($M = 2$) a pore state consisting of the **G,G'** microstates with a 50% population each (**Fig. 4a**) can be rendered. For comparison, **Fig. 4b** depicts in detail the previous 86-frame *indirect* EAP extracted from STM data (cf. **Fig. 2m**). Herein, the occupation probability for the microstate **G** is 0.44, which amounts to a combined **G,G'** occupation probability $p_{ref} = 0.88$ (i.e. the experimental value). To detail the analysis, a normalized free energy difference is defined as $\overline{\Delta F}_{A,B} \cdot T^{-1} = -R \ln (p_{ref\ B}/p_{ref\ A})$, where $R$ is the ideal gas constant (free energies are henceforth expressed in terms of $T^{-1}$ to show clear differences over a range of temperatures). Considering the state composed by microstates **G,G'** with respect to the EAP at 30 K, with $p_{ref\,A} = 1$ and $p_{ref\,B} = 0.88$, respectively, the normalized free energy difference amounts to $\overline{\Delta F}_{GG',EAP} \cdot T^{-1} = -R \ln(0.88/1) = 0.25$ cal mol$^{-1}$ K$^{-1}$.

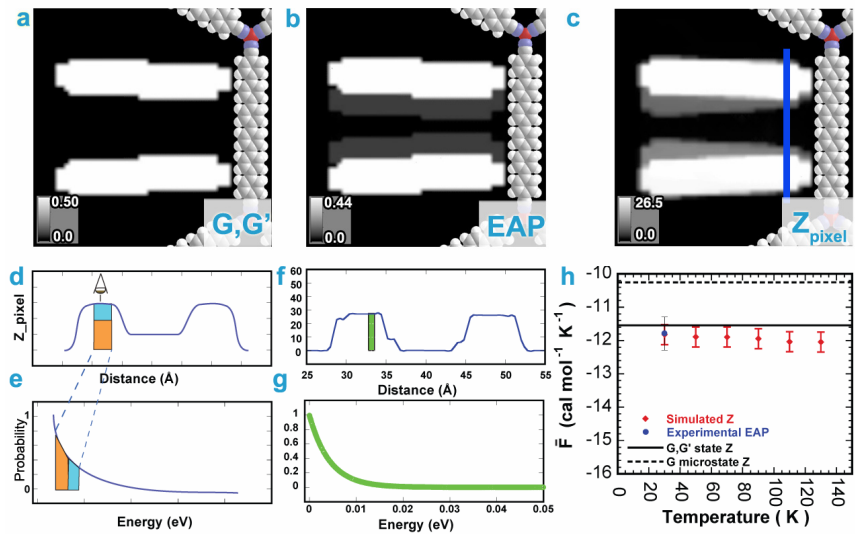

**Fig. 4. Images as quantitative thermodynamic measurements of a guided diffusion scenario.** **a)** Rendering of the static microstates **G,G'**. **b)** Ensemble-average (EAP) image rendering of 86 experimental molecular microstates at 30 K through the use of Equation 1. **c)** Sum over states (z) along pore rim at T=30 K, obtained by means of Equation 3, with M equaling 4x10$^4$ frames obtained from MD simulations (See Methods). **d,e)** Schematics of the configurational partition function per pixel in (e). **e)** Schematics exemplifying how two microstates with non-equal Boltzmann weights per pixel are summed to **z** of a single pixel in (d) within the current implementation of Equation 3 (see Methods). **f)** The value of a pixel at 30 K along the blue coordinate in (c). **g)** The Boltzmann distribution of the green pixel shown in (f). **h)** Normalized free energies from the sum over states and ensemble average patterns **a-c**.

Incidentally, we can obtain the normalized free energy from graphical real space analysis in order to establish quantitative correlations to molecular-level spatially-resolving experimental techniques. The normalized free energy $\overline{F} \cdot T^{-1}$ associated with an experimental STM TAP is obtained by summing over the image pixels with respect to the pixel reference value $p_{ref}$ of the considered microstate:

$$\overline{F_{ref}} = -RT \ln \sum_{n}^{no.pixels} \frac{p_n}{p_{ref}} \qquad (2)$$

As long as the number of pixels between two states is the same, **Equation 2** reduces to the normalized free energy difference introduced above. Using $p_{ref\,A} = 0.5$, for the a hypothetical reference state G',G representation in **Fig. 4a** this value is $\overline{F}_{GG'} \cdot T^{-1}$ = -11.54 cal mol$^{-1}$ K$^{-1}$. For **Fig. 4b** with $p_{ref\,B}$ = 0.44, $\overline{F}_{EAP} \cdot T^{-1}$ amounts to -11.78 cal mol$^{-1}$ K$^{-1}$. By taking the difference between the former and latter pore states we recover the free energy difference of $\overline{\Delta F}_{GG',EAP} \cdot T^{-1}$ = 0.25 cal mol$^{-1}$ K$^{-1}$, matching the value determined above. Note that the normalized free energy difference plays a convenient role: the closer it is to zero, the more the two compared pore states are energetically identical. Since $p_{ref}$ is also an occupation probability, **Equation 2** is only defined when $p_{ref} \neq 0$.

Finally, the visualization of the normalized free energy can also be generated from a microstate probability projected onto a real-space reaction coordinate. For this, we *project* (see SI) the configurational partition function in a pixel volume, $z_{pixel}$:

$$z_{pixel} = \sum_{i \in \{pixel\}}^{N} \exp^{-\beta U_i} = \sum_{k}^{MD-frames} \delta_{ik} \exp^{-\beta U_k} \qquad (3)$$

where $\beta = 1/kT$ and $U_i$ is the potential energy (**Suppl. Note 2**) of the *i*th-microstate configuration (with $3j$ atomic coordinates $x_1,...,z_j$) with respect to the G ground state of the whole system. The r.h.s. of Equation 3 shows that in order to extract the $N$ configurational microstates from a homogenous molecular dynamics sampling space (*MD*-frames), a delta function is used: delta ($\delta_{ik}$) equals 0 when the *i*th-microstate's configuration has been repeated and 1 otherwise. Note that **Equation 3** is an exact projection of the configurational partition function for three degrees of freedom if the $N$ microstate space is complete and the pixel volume is small enough to avoid spatial degeneracy.

In this work, for the single confined molecules in **Figures 4** and **5**, we use an effective pixel size of 1.0 Å$^3$ in a 80×80×8 Å$^3$ box size, forming a 51200 entry projected partition function matrix. A three-dimensional box/matrix is chosen in order to include all configurations (i.e. different interphenyl torsion angles, backbone bending, etc.). The size of the box was chosen for diagnosis and image rendering purposes, as well as to include the immobile MON frame. Using the r.h.s. of **Equation 3** with the number of *MD*-frames amounting to 5x10$^4$ as sampled by MD (see **Methods**), we construct the numerical partition function matrix at 30 K, rendering the position of the carbon atoms with pixels (**Fig. 4c**). The method is again easily understood by tracing a blue line across the so-constructed projected partition function matrix in **Fig. 4c**. First, a single pixel $z_{pixel}$ is highlighted across the profile, as illustrated with a rectangular box in **Fig. 4d**. Our computational implementation of **Equation 3** (see **Methods**) makes the value of a single pixel correspond to a sum over Boltzmann weights as depicted in **Fig. 4e**. Actual values of **Fig. 4c** are highlighted in green in **Fig. 4f,g**. Note that our computational implementation of integrating over energy levels, rather than in configurational states, disregards the density of states contribution due to spatial overlap of degenerate states, and is discussed below.

Finally, the projected sum over states matrix can be further reduced to a simple 2D image in order to compare to the 2D imaging experiments. In this graphical form, **Equation 2** can be applied to **Fig. 4c** (using $p_{ref}$ as the value corresponding to the area of the molecule, see **Suppl. Note 3**) leading to $\overline{F}_z \cdot T^{-1}$ = -11.86 cal mol$^{-1}$ K$^{-1}$. The difference with the **G**,**G'** state, $\overline{\Delta F}_{G'G,z} \cdot T^{-1}$ = 0.32 cal mol$^{-1}$ K$^{-1}$, is in agreement with the value calculated from the indirect ensemble-averaged pattern (EAP), $\overline{\Delta F}_{GG',EAP} \cdot T^{-1}$ = 0.25 cal mol$^{-1}$ K$^{-1}$. The difference between model ($z$) and experiment (EAP) at this stage is below 0.1 cal mol$^{-1}$ K$^{-1}$. Such image free energy comparison is expected to complement current structure-model correlations by, e.g. root mean square atomic deviations. It is worth mentioning that the excellent correlation between experiment and modeling is also surprising: because **Equation 3** is not an average, it takes only one microstate in the pore (out of hundreds of thousands of microstates) with an energy deviation of less than 20 meV (i.e. the energy difference between G and I$_2$), to completely disagree with the experiment. Most importantly, given a set of *MD*-frames sampled at a specific temperature, our method can be used to calculate the normalized free energy for all points in temperature space below the MD sampling temperature (**Fig. 4h**). The normalized free energy's absolute minimum for our one molecule system $\overline{F}_G \cdot T^{-1}$ = -10.15 cal mol$^{-1}$ K$^{-1}$ is also shown, which is a measure of the minimum pixel occupation of one molecule. With **Equation 3** and the aforementioned examples at hand, it becomes clear that at the ergodic limit, a complete energy microstate space in the pore system can be sampled for the temperature of choice. At this point, the sum over states $z_{pixel}$ becomes in fact, a *projected* configurational partition function.

## 6 Direct STM TAP as finite-temperature *projected* partition function (PPF)

Because a direct TAP in STM through the confinement approach satisfies ergodicity, an equilibrium normalized free energy can be correctly and sub-molecularly measured at experimental temperatures. Moreover, with the numerical method for the sum over states, a direct correlation between the STM TAP imaging and the corresponding calculated PPF modeling is possible. **Fig. 5a** shows the STM TAP image at 82±2 K, whereby the measured corrugation amplitudes are normalized to the occupation probabilities of black (0) white (1), respectively (see **Methods**). The data nicely correlate with the PPF modeling at T=80 K (**Fig. 5c**), constructed using again *MD*-frames sampling by µs-long Langevin MD simulations (See **Methods**, note that STM TAP images are now compared to PPFs further reduced to two dimensions). The blue arrows in the experimental and simulated images also show how it is possible to even reproduce geometrical features with marginal occupation probabilities, which we attribute to a diagonal adsorption microstate (see **Fig. S8** e,j). Note how sub-molecular geometrical features present in the high-resolution STM TAP image can be reproduced in great detail when using smaller matrix pixel sizes of 0.5 Å and fitting the atoms with Gaussian envelopes (see triangle in inset **Fig. 5a,c**; such resolution is computationally very expensive and its broad application not practicable).

Interestingly, the PPF in **Fig. 5c** appears to overestimate the experimental TAP occupation probabilities near the hexagon's center. This is a first indication that our computational implementation of **Equation 3** (**See Methods**) does not significantly omit degeneracy due to symmetry crossing of G states (see increased probability near the pore center in **Fig. S10a**). We ascribe this to an error compensation effect: because of a finite numerical precision, two microstates that should be in principle identical, rarely have exactly the same energy. Instead, we attribute the discrepancy to decreased sampling near the pore edges. Notably the intrinsically limited thermalization of the simulated hexagonal nanomesh (see **Methods**) artificially increases the amount of states sampled in the pore center. This is shown in **Fig. 5g**, revealing that the temperature near the rim is reduced (T ≈ 223 K) from the original configuration space sampled at 250 K. We then examined the STM data at the experimental maximum temperature of 145 K (**Fig. 5b**; upon exceeding this temperature the guest species leave the pores and move across the entire nanomesh), implying orders of magnitude higher diffusion rates compared to the previous 82 K situation. Under such conditions, the main TAP features are retained along with the reduced occupation probability at the pore center. Normalized free energies can be extracted by means of Equation 2. For this, we use $p_{ref}$ = max($p$), setting the occupation probability of the most probable *pixel* state to 1 at finite temperature in the pore (**see Supp. Note 3**). **Fig. 5e** shows now the quantitative correlation of the PPF vs. direct TAP values at T=80 and 145 K, respectively. For instance, at 80 K the difference is as small as 0.2 cal mol$^{-1}$ K$^{-1}$. Note how the normalized occupation probability, and thus the free energy difference, between pixel states near the G state and the C state is reduced with increasing temperature. The inset of **Fig. 5e** shows the plot of the free energy vs. 1/T between 100 K and 250 K, where the pixel entropy and pixel enthalpy of the PPF method can be estimated. The entropy amounts to $S_{PPF}$ = 14.3 ± 0.4 cal mol$^{-1}$ K$^{-1}$ and $H_{PPF}$= 67±5 cal mol$^{-1}$.

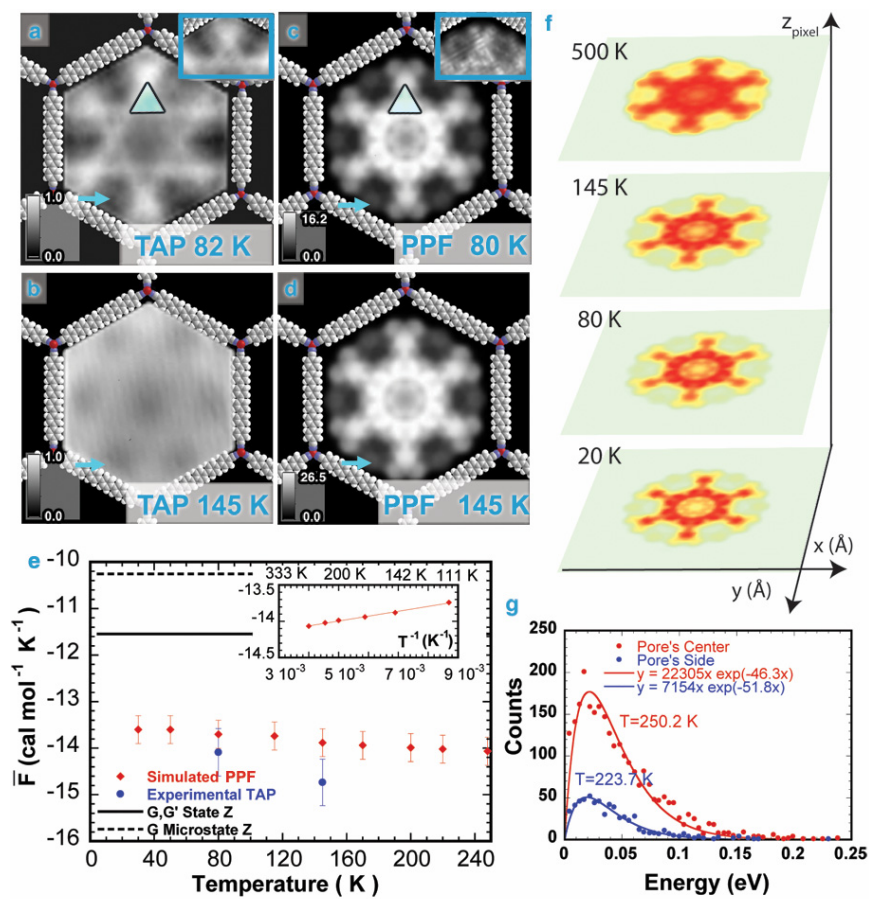

**Fig. 5. Time-averaged patterns and their modeling. a,b)** High-resolution STM time-averaged image at 82±2 K and 145±5 K respectively. Tunneling parameters $V_B$=5 mV, $I_t$ = 100 pA. Blue triangle side 9.5 Å. **c,d)** Projected partition function (PPF) renderings of a single molecule in the pore at 80 and 145 K. **e)** Differences between TAP and the PPF through normalized free energy of a single confined molecule at different temperatures computed using the r.h.s. of Equation 3 and Equation 2. The normalization is carried out using $p_{ref}$ as the maximum value in the confined pore. The inset shows the linear fit F ($T^{-1}$) = -14.3 + 67 $T^{-1}$ using data between 100 K and 250 K **f)** Changes in the 2D pattern of the projected partition functions vs. temperature. Images were rendered using carbon pixel projections in the aromatic backbone only with a pixel size of 1 Å$^3$. Free energies were extracted from similar images rendered by decreasing their resolution to 80 x 80 pixels$^2$. **g)** The molecular velocity distribution in MD simulations at 250 K of single 1 Å$^2$ pixels in the center of the pore and near the rim. Maxwell-Boltzmann fits show the respective pixel temperatures.

**Figure 5f** summarizes the modulation of the PPF over a wide temperature range. The chosen color-coding emphasizes the details between the PPFs at 20, 80 and 145 K, which arise due to the increasing occupation of a manifold of microstates. These findings demonstrate the evolution of distinct patterns in (*x,y,T*)-space at the ergodic limit. **Figure 5f** also includes a hypothesized PPF modeled at 500 K where a purely entropic situation (homogeneous pore sampling) might be expected. However, the PPF retains an articulated pattern near the pore's center, exemplifying the aforementioned temperature gradient.

Next we demonstrate how to systematically design the potential energy landscape for encoding information in ($x,y,T$)-space, which approach extends and complements the holographic encoding schemes reported previously exploiting 2D fermionic states tailored by molecular manipulation[40]. The specific example chosen is the enthalpy-driven expression of the letters *I*, *L* and *U* (**Fig. 6a-c**). The nanopore's potential energy landscape is modulated by oxidizing two cobalts in the rim and reducing a molecule (see **Methods**). In this configuration, a single ground-state adjacent to the pore's rim is favored, thereby providing the letter *I* at very low temperatures. At intermediate temperatures, the molecule occupies with significant probability its second most favorable state by binding to the oxidized cobalt coordination node, thereby tracing an *L*-like configuration. Finally, at higher temperatures, additionally population of a third microstate valley around a configuration pointing towards a second oxidized cobalt results in an *U*-shaped pattern. **Figure 6d-e** depicts the simulated PPF images demonstrating the simple enthalpy-driven scheme. By sequentially changing the PPF temperature from 20 to 1 to 400 K the encrypted message reads "*L I U*", acronym of Linköping University. Though the realization of the corresponding experimental system would require further major efforts, it can be concluded from the modeling alone that the PPF method acts as a powerful local-temperature probe or alternatively as sensor allowing fast prototyping of thermodynamic encryption and decryption schemes.

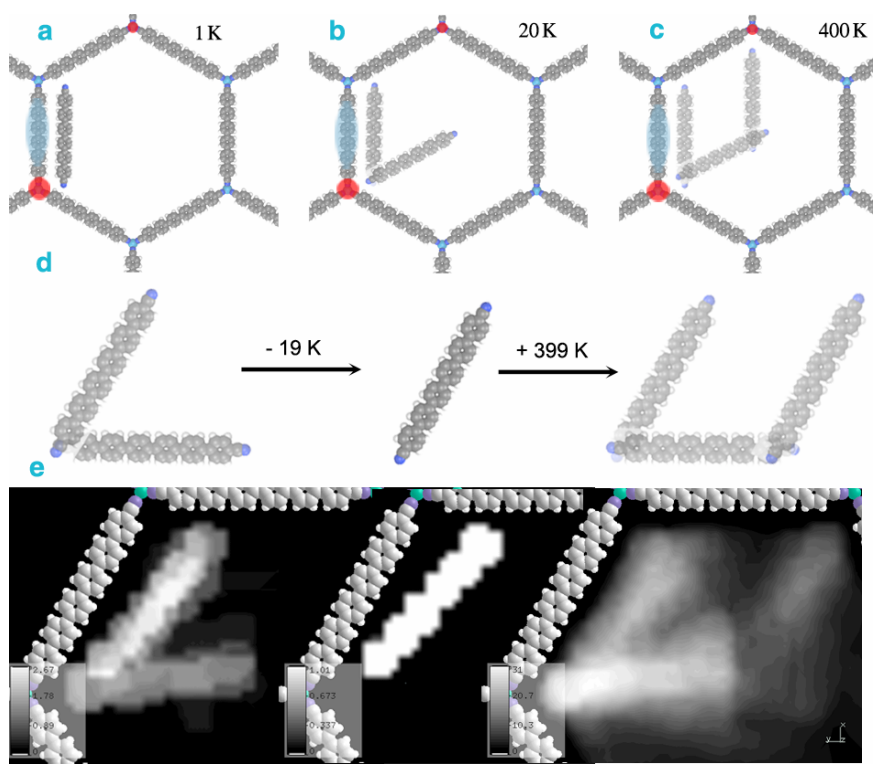

**Fig. 6. Simulated thermodynamic encoding of the acronym L I U. a-c)** The energy landscape of the grid is modified to favor a single non-degenerate *I* ground state at low temperatures, a degenerate *L* average at medium temperatures, and a *U*-shape at high temperatures. The landscape is steered by reducing a single rim molecule adding approx. 0.7 negative charge (blue oval) and oxidizing the cobalt coordination spheres (red circles). **d)** The acronym L I U can be decrypted by changing the temperature from 20 K to 1 K and then to 400 K, **e)** Projected partition functions at 20, 1 and 400 K from an $M=1\times10^4$ sampling space.

## 6 Summary

We have established a quantitative link between visualizations of simulated projected spatial partition functions and experimental STM time-averaged patterns in confined spaces and at different temperatures. Because of the confining nature of the employed single-molecule environment, both experiment and simulation represent thermal equilibrium whence single-molecule ergodicity applies. This allows for visualizations of thermodynamic probability distributions in position-temperature space, opening up the field of thermodynamic information encoding. We expect rapid advances in custom-design systems yielding informed thermodynamic patterns through a varying temperature-dependent signature. Our investigations suggest the realization of nanoscopic thermal probe/sensors that thermodynamically encrypt/decrypt information. Apart from such emerging fields, our observations in confined spaces, theories and methods have immediate applications in the study of phase transitions, ergodicity breaking and analytical development of force fields and density functionals. We foresee that further simulations and observations of projected partition functions will continue with the quantitative elucidation of molecular interactions and reactions at finite temperatures in real-time, with sub-nm spatial resolution.

## Methods

**Sample preparation and scanning tunneling microscopy.** All experiments were performed under ultrahigh vacuum conditions using a home-built liquid-He-cooled low-temperature STM with cryoshields. The employed Ag(111) substrate was prepared by cycles of Argon sputtering and annealing. Sexiphenyl dicarbonitrile molecules were deposited from a quartz crucible in an molecular beam source at 572 K with the substrate was kept at 300 K. Sub-monolayer molecular films were subsequently exposed to a beam of Co atoms to induce the metal-directed assembly of a MON. Further molecules were added at 144 K at distinct sub-monolayer coverages to obtain a preferential decoration of honeyomb pores by monomers. All the STM images shown were recorded with ~ 4Hz frequency in fast-scanning direction and subject to line-wise leveling. Data presented in this work exclusively show nanopores in which a single molecule is trapped.

**Density functional theory.** The periodic DFT calculations were performed with the VASP code[44], with the ion-core interactions described by the projector-augmented wave method[45]. The PBE functional[46] was used, together with the Grimme correction[47] to include van der Waals interactions. Note, that the Grimme correction was not included between Ag atoms, in order to avoid unphysical shrinking of the Ag slab. If not indicated otherwise, the calculations were done with a 400 eV plane wave cutoff, and the 1$^{st}$ Brillouin zone was sampled by the Gamma-point only. A slab of three layers represented the silver surface. In the structural optimizations, the atoms in the molecules, Co-adatoms, and the atoms in the outermost Ag surface layer, were allowed to relax until the residual forces acting on each atom were smaller than 0.03 eV/Å.

**Molecular dynamics.** All-atom MD simulations were performed with the CHARMM 36b2 Package[48] using the scripts provided in the SI. The simulation system featured 7811 atoms (carbon, nitrogen, hydrogen, silver and cobalt atoms) in an isolated nanopore with infinite non-bonded cut-offs. Langevin and Nose-Hoover thermostats were used with three different thermostat friction coefficients and integration timesteps of 2 fs. The SHAKE[49] module was used to constrain all C-C and H-C bond lengths. The C22 parameters were used for providing C and H bending, dihedral and vdW parameters during MD simulations. Thus, only some vibrational, rotational and translational degrees of freedom are considered in our system. All electrostatics and adsorption vdW parameters were parameterized following procedures shown in the SI. The metal substrate and the metal-organic framework pore were kept fixed during the simulations. For sampling the microstates bound to a honeycomb segment in Fig. 4, Langevin[48] MD simulations with a friction of 0.01 ps$^{-1}$ were performed at a temperature of 100 K. This temperature is chosen in order to allow the molecule to diffuse from the microstate in **G** to the intermediate ones but not to adjacent pore sides on the microsecond time scale. For sampling the whole pore in Fig. 5, Langevin[48] MD simulations were recorded at higher temperatures (T=250 K), as means to achieve homogeneous sampling of the pore at the simulation time scales. To obtain the *MD*-frames in Equation 3, three independent 600 ns trajectories with integration steps of 0.002 ps were combined, at temperatures given on the text. The convergence for the 100 K trajectories as a function of the image free energy is given in Figure S9.

**Projected partition function method.** The PPF formalism is developed in the SI. For computational efficiency, the r.h.s. **Equation 3** was numerically integrated in energy (levels), rather than in configurational (states) space. This computational implementation removed degeneracy per pixel with precisions of 1 meV pixel$^{-1}$ (~23 cal mol$^{-1}$ pixel$^{-1}$ ). this approximation is equivalent to excluding the density of states during the sum over energy. Because the density of states is not explicit in this implementation, in special cases it ignores energy degeneracy, as discussed in the SI. Detailed information on this computation and scripts are provided in the SI. Uncertainties in the normalized free energies from the PPF and indirect TAP and EAP methods were computed from the standard deviations for pixel size renderings between 1, 0.9 and 0.8 Å. For the direct TAP from experimental STM images, the standard deviations are computed allowing 20% changes in the images contrast. For the TAP analysis of STM images in Fig. 5a,b a Gaussian filter was first applied, followed by cropping to 80 Å x 80 Å. The area not belonging to molecular confinement was removed with an hexagonal mask and replaced by superposed models of the honeycomb pores. The empty area's color code was adjusted to the color of one adjacent empty pore in the same STM image and the whole picture subsequently equalized until the empty area featured a color black with an ASCII value of zero. All direct TAP and PPF images shown in the main text are rendered with a linear extrapolation made with the program V_SIM of the CNRS, CEA and INRIA http://www-drfmc.cea.fr/L_Sim. Except for Fig. 4a-c where magnification and color levels were adapted for clarity, the free energies were extracted directly with Equation 2 from Fig. 5 images as-rendered and presented in this work (in ASCII file format with a resolution of 80 x 80 pixels, excluding superposed frame models and by means of the potsummed.py script in the SI). For the thermodynamic encryption in Fig. 6, an MD simulation was performed at 400 K with the same parameters except the charge in two parameterized cobalt atoms (i.e. the coordination sphere of the cobalt) was set to 0.15. A third cobalt between them was also slightly oxidized to a charge of 0.025. The two central carbons of a neighbor molecule were assigned a charge of 0.35 each. The PPF analysis was then performed with the same charge modifications at the temperatures described in the text.

**Supplementary Note 1.** For sake of MD sampling efficiency, the electrostatic parameters are tuned to underestimate the DFT barriers between the microstates, while reproducing the microstate energy differences (**Fig. 3d**, green line). Despite of this approximation, we find that the rotation rates of a caged trimeric dynamer in a pore with our model at 250 K of (5.6 ± 2.6) x $10^7$ Hz, are in good agreement with experimental rotation rates of 1.8 x $10^8$ Hz (data extrapolated from [28]).

**Supplementary Note 2.** We note that although geometrical minima in the MD force field is reproduced with great level of detail (**Fig. S7**), by parameterization, only the electrostatic energy without vdW contributions reproduces the correct PBE interaction energy (see **Fig. S8**). In fact, while we have previously found that the CHARMM C22 Lennard-Jones (LJ) parameters describe remarkably well C-C and C-H interactions when charges are not present[50], in the presence of the aforementioned increased charge parameterization, the LJ parameters and function itself underestimate the energetic repulsion ($-C_{12}/r^{12}$) quite considerably[51]. Therefore, throughout this work including **Equation 2**, the reported force-field energies correspond to the electrostatic energies plus the hydrogens' CHARMM LJ interactions only. This minimal LJ interaction is needed in order to avoid overestimating the electrostatic energetic contribution of structures that are repulsive (i.e. close atomic overlap).

**Supplementary Note 3.** Note that the normalized (image) free energy is an independent sum of pixel probabilities (**Equation 2**) and is valid only when using pixel-wise (reference) states instead of configurational, molecular ones (see **Fig S10**). The use of $p_{ref}$=max(p) is made to reference the normalized free energy to the most probable pixel-state. **Equation 2** can also be used to compare molecular rather than pixel states under certain conditions, as explained previously: When using $p_{ref}$=1/2 (in **Figure 4b** a value exclusive to the G' molecular microstate area) the free energy of the G'G molecular state (F = -11.54 cal mol$^{-1}$ K$^{-1}$) is closely recovered by adding the degeneracy of -Rln(2) to the G' molecular state (i.e. F = -10.15 cal mol$^{-1}$ K$^{-1}$ - Rln(2) = -11.52).

## Acknowledgements

Work supported by the European Research Council via Advanced Grant MolArt (Grant 247299). J.B. and S.S. acknowledge the Swedish Research Council for funding. All authors warmly thank Mario Ruben and Svetlana Klyatskaya for providing the molecules used in the experiments, and Marie-Laure Bocquet and Vladimir García Morales for helpful comments. C.-A.P. thanks Martin Spichty for useful discussions. We thank Adolfo Poma and Simon Poblete for further improvements, and the Swedish National Supercomputing Center project SNIC 001/12-83 for allocated supercomputing resources.

# *Annex: Supplementary Information*

## *Simulation and Methods Table of Contents*

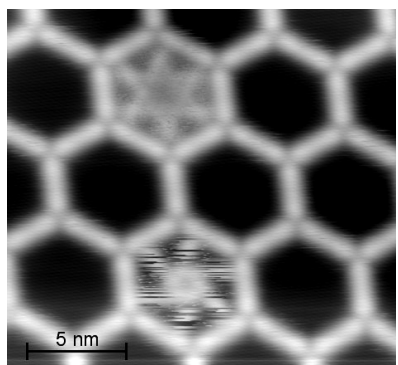

**Figure S0.** STM image of the nanomesh with simultaneous imaging of caged monomers and trimers at T ~ 80 K. The hexagonal pattern created by the rapid monomer motions in the upper part is clearly distinct from the 2D chiral signature of the supramolecular trimer in the lower part, performing the rotational motions described in Proc. Natl. Acad. Sci. U.S.A

# 1) Electrostatic Potential Parameterization

## Full scripts given in the online attachment *Elec_Pot_Tools_All_SI.tar.gz*

The general strategy of potential parameterization from DFT to point charges (electrostatic force field) consists of **a)** Automatic 3D Potential parameterization via point charge random walk **b)** Point charge refinement via comparison to 1D electrostatic potential plots potential at z = 3Å at different x,y positions (pore only) **c)** Point charge refinement via comparison to 1D energy plots of pore-molecule interaction at z = 3Å and x = 2.18 Å for the ground microstate (**Figure S1**).

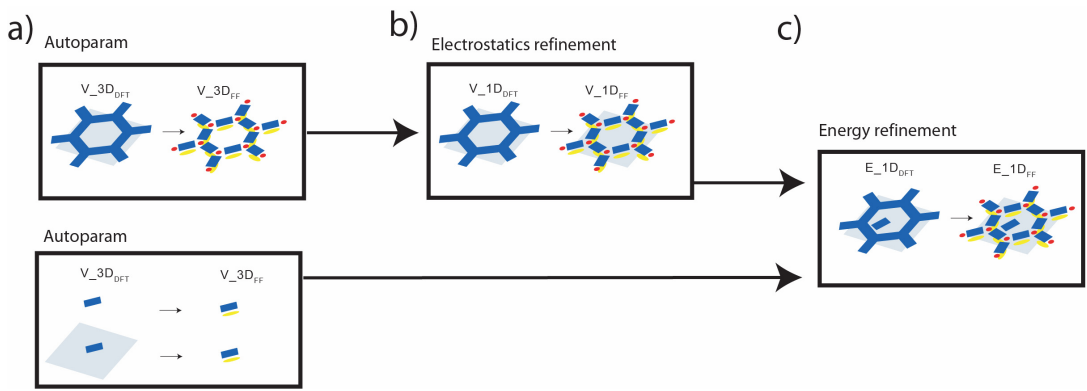

**Figure S1.** General strategy for fitting the DFT electrostatic potential ($V_{DFT}$) and DFT interactions ($E_{DFT}$) to a molecular force field ($E_{FF}$): **a)** Automatic 3D Potential fitting via point charge random walk **b)** Point charge refinement via comparison to 1D potential plots **c)** Point charge refinement via comparison to 1D energy plots. The yellow highlights indicate additional "dummy atoms" used for image charges. The cyan square indicates the substrate.

### a) Electrostatic potential extraction (density2elec folder)

i) When needed, the electronic density of the DFT relaxed structure (a VASP *CHGCAR* file) is first truncated to a lower resolution. The electronic density is transformed to ascii format.
*Command: python2.7 CAR2dens.py CHGCAR (writes CHGCAR.pot)*
An ascii coordinate file is extracted from the CHGCAR with the number of valence electrons per atom. *(writes CHGCAR.ascii)*

ii) An FFT poisson solver is then used to construct the corresponding electrostatic potential in a 3D. ( *dens2pot.py*).
*Command: python2.7 dens2pot.py CHGCAR.pot CHGCAR.ascii (writes CHGCAR_pot.pot)*

iii) The electrostatic potential is then coarsed from ca. 0.07pixels per Ångström to ca. 0.3. (**Figure S2 and S3a**)
*Command: python2.7 reduce_split_pot.py 4 CHGCAR_pot.pot (writes CHGCAR_pot_coarse.pot)*

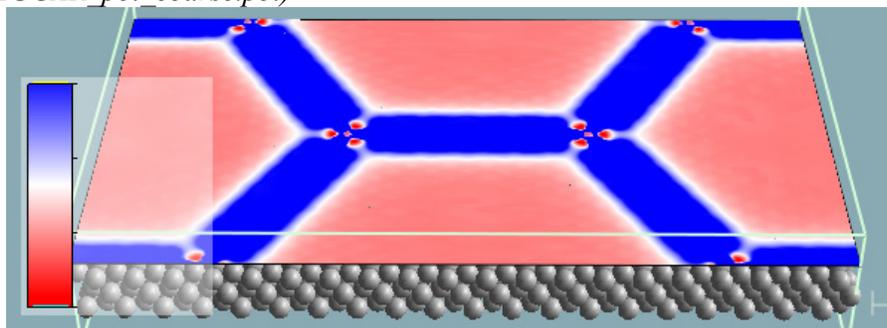

**Figure S2**. Electrostatic potential slice at 3Å from the surface, between -500 and 500 mV for the full DFT pore's unit cell.

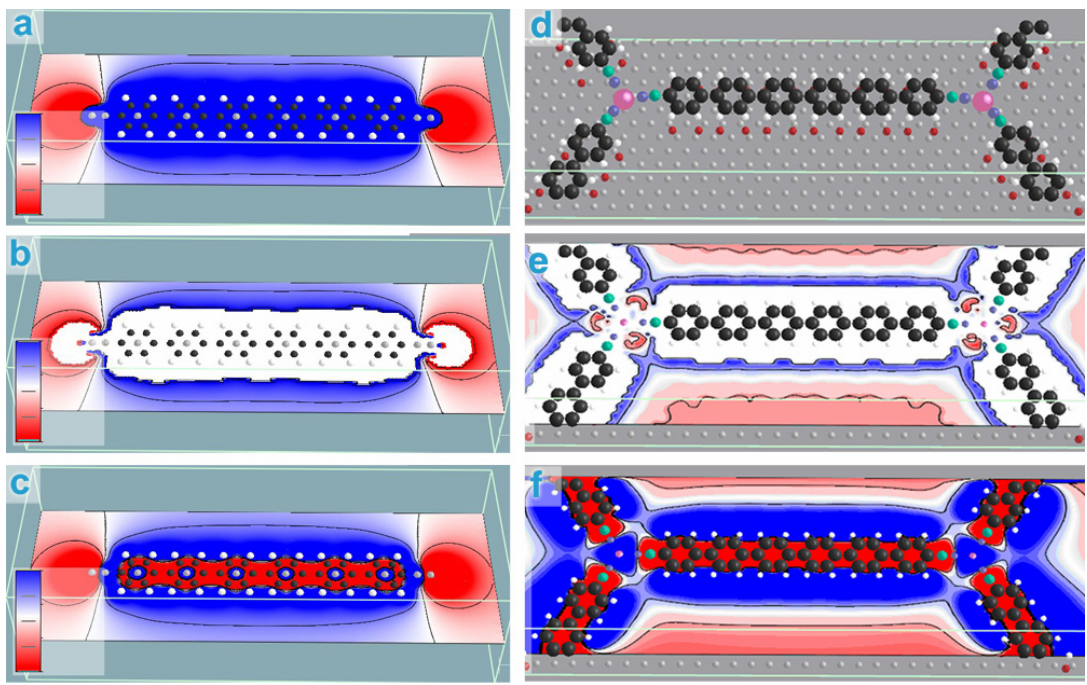

**Figure S3**. **Electrostatic potentials fitting of the sexiphenyl dicarbonitrile in the pore and the pore.**
Electrostatic potential slice of the sexiphenyl dicarbonitrile at the molecular plane between -500 and 500 mV in sexiphenyl dicarbonitrile for **a)** the full DFT unit cell and **b)** the masked DFT unit cell (cutting everything between -500 and 500 mV) and **c)** the autoparameterized unit cell. **d)** The chosen atoms for parameterizing the truncated pore unit cell in Figure S2. The point charge elements used for the automatic fitting are Black: Carbons. Cyan: CN Carbon. Blue: Nitrogen. Magenta: Cobalt. Red: Image dummy atoms of the hydrogens. Gray atoms represent and ad-hoc construction of the Ag(111) consisting of 3 merged slabs of silver Ag(111) and are not used in the automatic fitting procedure. Throughout this work at all times, the positions of all elements are given by the same structure obtained from PBE-DFT minimization. Comparison between slices at 3 Å from the surface from **e)** the masked DFT electrostatic potential and **f)** the potential as-obtained from the random walk over a space of Hydrogen, Carbons, Cobalt and Dummy point charges. Neither nitrogens nor surface atoms are considered in this particular fit and the CN carbon atom is grouped as a carbon-atom type. (See **Table S1** for autoparameterized charges ). The color scale represents the potential between -500 and 500 mV.

### b) *Autoparametrization through Random Walk (autoparam folder)*

i) Once the electrostatic potential from the DFT electron density is extracted, an all-atom structure is created for the point charge autoparameterization. For sexiphenyl dicarbonitrile alone, this is shown superposed with the DFT potential in **Figure S3a.** For the hexagonal pore, this atom file is shown in **Figure S3d**.
ii) For the autoparameterization, the potentials higher or below than ±500 mV are masked as shown in **Figure S3b** and **Figure S3e**.
iii) Only 4 atom types are chosen for a random walk in charge space. (see **Table S1** and **Table S2**, where n/a indicates atoms left out of the autoparameterization)
iv) The masked potentials (from point charges and DFT) are compared as obtained through a random walk in charge space ( *random_walk.py* ).
*Command: chargeset.x (writes CHGCAR_(initialcharges)_(finalcharges) _(masked_DFT_E_density-masked_parametrized_E_density).pot*
v) Because of computational cost, the potential of **Figure S2** is further truncated and its resolution decreased to a grid mesh of 0.46 Å as shown in **Figure S3e.**
vi) The electrostatic potentials with the lowest difference between the masked DFT reference and the masked point charges are chosen as a starting point for refinement and shown in **Figure S3c** and **Figure S3f**.

c) *Electrostatic Potential Refinement (ascii2elec folder)*

Because the substrate charges cannot be successfully treated in our random walk autoparameterization, these have to be manually tuned. For this end, a plot of a slice of the electrostatic potential is obtained, at a height of z= 3.0Å and at 2.0 Å from the nearest pore hydrogen (**Figure S4**), in a direction parallel to the pore's side, as shown in the inset of **Figure S4.** The plot is compared to the same slice in a fully-hexagonal all-atom point-charge representation of the pore. The final manually-refined charges matching the DFT electrostatic potential and including each of a 3-slab silver substrate atoms is shown in the black curve, **Figure S4**.

> *Command: ascii2Epot.py CHGCAR.ascii (writes CHGCAR_(finalcharges) .pot) – where, the charges in the ascii2Epot.py are continuously manually tuned departing from the autoparam charges.*

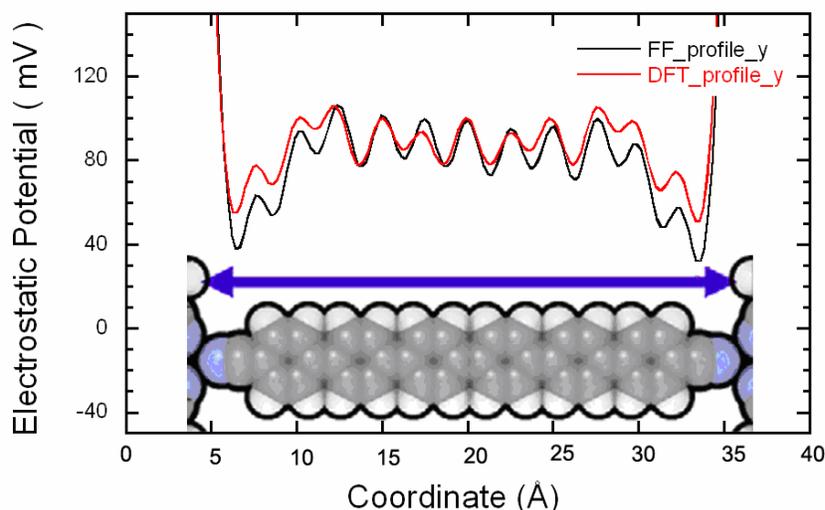

**Figure S4.** Electrostatic Potential refinement of the Force Field (FF) point-charge model to the PBE-DFT taken at a profile of at a height of z= 3.0Å and at 2.0 Å from the nearest pore hydrogen.

d) *Potential Energy Interaction Refinement*

So far we have managed to obtain parameters describing atomic charges with an accuracy better than 10 mV/Å. Of particular interest is to verify the energies arising from the FF parameterization to those obtained directly with PBE-DFT. For obtaining a near-perfect match of both FF and PBE-DFT interaction as shown in **Figure 3b** main text green curve, the point charges of both the sexiphenyl dicarbonitrile and the pore's hexagon had to be increased as shown in **Table S1** and **S2**. This increased charge parameterization is naturally done in order to compensate for non-electrostatic, non-dispersive and polarization effects of the interaction energies in silver.

For sexiphenyl for instance, this corresponds to the an increased electrostatic potential as observed in **Figure S5a**. For the pore's hexagon, this corresponds to ca. 50 meV electrostatic potential increase around the slice shown in Fig. S4 (**Figure S5b,** black line).

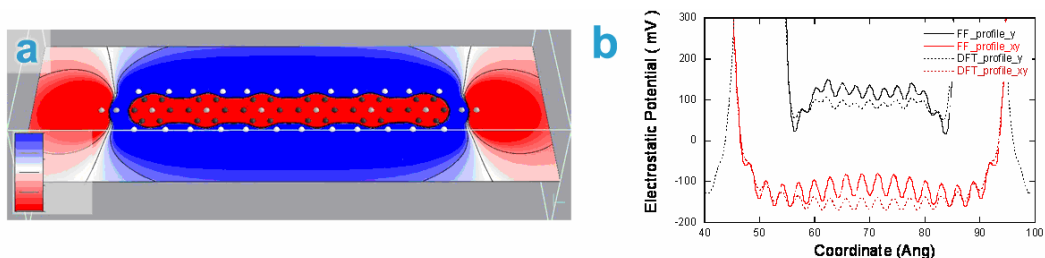

**Figure S5.** Increased Electrostatic Potentials for matching interaction Energies as the final FF parameterizations used in the current work. Increased point-charge electrostatic potentials of **a)** a single sexiphenyl dicarbonitrile and **b)** the pore side hexagon: The slice as shown in Fig. S4 (black line) and the slice at the same height but tracing a diagonal through out the pore (red). The color scale represents the potential between -500 and 500 mV.

In the last step, Lennard Jones (LJ) parameters are obtained for the silver atom and fitted to the PBE-DFT-D adsoption curves of one sexiphenyl ontop of a periodic silver slab (**Figure S6**).

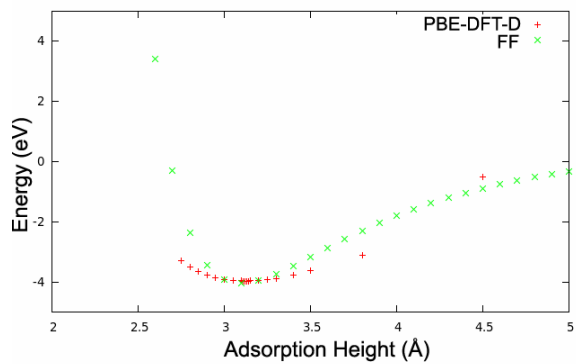

**Figure S6.** Parameterization of the adsorption of sexiphenyl dicarbonitrile on silver, using the final refined charges and Lennard-Jones parameters.

Finally, the equilibrium distances and energies (rather than the scans along one direction shown in **Fig. 3d** main text) are matched to those of PBE-DFT-D by tuning the Lennard-Jones (LJ) parameters in our force field (FF). The equilibrium geometry of the FF accurately reproduces the PBE-DFT-D with the exception that it is 0.5 Å tilted away from the silver hollow site (**Figure S7b**). This is explained in a straight-forward way. As introduced in the main text, silver top sites are more neutral (hollow sites are strongly negatively charged). As observed in **Figure S3a**, the center of the nitrogen itself is positively charged and it is the surrounding area which is negatively charged. In the FF however, mimicking this anisotropic charge distribution is difficult without the use of additional dummy/point charges. Because of this we parameterize nitrogen only with one negative charge (**Figure S3c**) and as a result, nitrogen does not sit favorably on top of the negative silver hollow site.

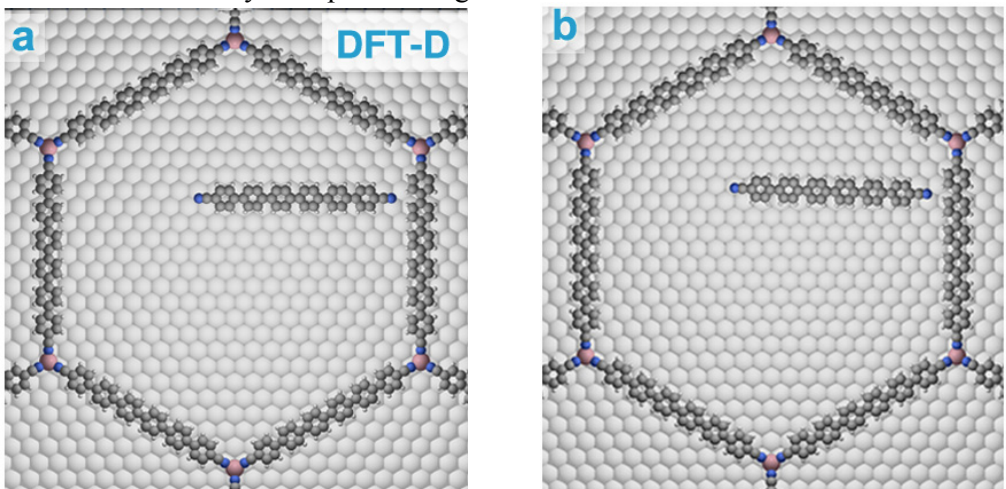

**Figure S7**. The geometry of the G microstate in **a)** DFT and **b)** FF calculations.

In non-bonded interaction energetic terms, our force field is an electrostatic charge only force field. Because of this the best agreement with the PBE-DFT energies are found when the force field does not include carbon nor nitrogen LJ parameters, rather hydrogen-hydrogen LJ parameters alone (**Figure S8**). Of course, for actual MD sampling, carbon and nitrogen parameters are needed in order to keep the molecule horizontal to the substrate. Because of this reason, simulations sampling the *MD-frames* are performed with native C22 Charmm LJ parameters, while the PPF method only considers the hydrogen-hydrogen LJ parameters. We rationalize this incompatibility as a limitation using LJ functional form for all atom types, instead of independently tuning the damping functional form for each atom (metal, hydrogen, carbon).

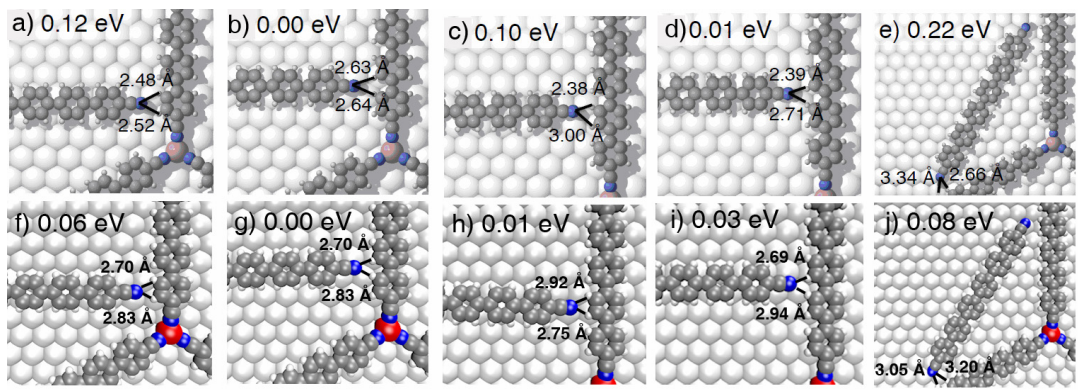

**Figure S8**. Comparison between **a-e)** PBE-DFT energies computed through minimized PBE-DFT –D corrected calculations and **f-j)** electrostatic Force Field energies computed through Elec-vdW Force Field calculation.

**Table S1. Summary of Pore electrostatic** point charges parameters for the three refinement steps above.

| Element Name | a) Autoparam | b) Electrostatic Potential Refinement | c) Energy Refinment |
|---|---|---|---|
| H | 0.25 | 0.1500 | 0.2000 |
| C | -0.10 | 0.0100 | -0.0750 |
| C(N) | -0.10 | 0.0620 | 0.1125 |
| N(C) | n/a | 0.0000 | -0.0375 |
| Co | 0.20 | -0.1500 | -0.2250 |
| D | n/a | -0.1515 | -0.1107 |
| Ag(M) | n/a | 0.6000 | 0.8000 |
| Ag(W) | n/a | -0.3000 | -0.4000 |

**Table S2. Summary of sexiphenyl dicarbonitrile electrostatic** point charges parameters for the three refinement steps above.

| Element Name | a) Autoparam without surface | c) Energy Refinment |
|---|---|---|
| H | 0.1 | 0.24 |
| C | -0.12 | -0.12 |
| C(N) | 0.34 | 0.7 |
| N(C) | -0.4 | -0.8 |
| Co | n/a | n/a |
| D | 0.34 | -0.20666 |
| Ag(M) | n/a | n/a |
| Ag(W) | n/a | n/a |

## 2) Configurational Projected Partition Function Formalism

Direct extraction of a partition function from an MD simulation is not possible, because the volume of phase space cannot be measured directly. Therefore ideal-gas approximations ( see D. R. Herschbach, H. S. Johnston, and D. Rapp, J. Chem. Phys. **31**, 1652 (1959) ) and projections (or coarse-graining, CG e.g. W.G. Noid, J. Chem. Phys. **139**, 090901 (2013)) are needed in order to reproduce from the bottom-up ensemble and time-averages obtained from the top-down.

The classical configurational partition function in Cartesian space is,

$$Z = \int ... \int \exp(-\beta U) \, dx_1 ... dz_j$$ **Equation S1**

Where *j* are the number of atoms and U is the potential energy for a specific configuration $x_1,...,z_j$.

Importantly, the configurational partition functions omits the integration over the momentum degrees of freedom, which cancel out when taking an isothermal free energy differences between the same species (D. Chandler, *Introduction to Modern Statistical Mechanics*. Oxford University Press, USA, 1st ed. (1987)). It is worth noting that the potential U does not consider excited electronic states. Another important consideration is that the potential used stems from a constrained molecular dynamics simulation (See main text Methods). This implies that only some vibrational, rotational and translational degrees of freedom are considered in our integration of the configurational partition function.

The configurational partition function can also be expressed as,

$$Z = \sum_{i}^{N} \exp(-\beta U_i) \quad \text{Equation S2}$$

provided that *i* corresponds to a uniquely defined configuration and that the *N* configurational space is complete.

**Equation S1** can be projected in a local partition function coordinate *V(x,y,z)*,

$$Z(V) = \int ... \int \exp(-\beta U) \delta\big(V(x_1,...,z_j) - (x_1,...,z_j)\big) dx_1...dz_j \quad \text{Equation S3}$$

This function can be coarse-grained and integrated numerically via Equation S2,

$$Z(\Delta V) = \sum_{i \in \{x_1,...,z_j\}}^{N} \exp(-\beta U_i) \, \delta_i(\Delta V) \quad \text{Equation S4}$$

where the delta function equals 0 if the coordinates of the i*th* configuration do not belong to the volume element $\Delta V$ and 1 otherwise. Equation S4 is nothing more than $z_{pixel}$ in **Equation 3** in in the main text, where the $\delta_i(\Delta V)$ has been omitted for brevity.

## 3) Numerical Computation of the Configurational PPF

## Full scripts in ( *pynalysis_energy.tar* )

### a) Provoking Ergodicity

Molecular dynamics sampling of the complete configurational energy hypersurface is a colossal challenge even for the model nanoscopic single-molecule system presented in this study. As stated in the main text, we use different strategies for the numerical integration of each possible Boltzmann weight correctly describing a complete full partition function (**Equation 3** main text) from the configurational (*N*-space) arising from the MD sampling (*MD-frames*).

*i. Reduction of the MD-frames to a 2D equilibrium position and symmetrization*
First, we aid the convergence of the *N*-space by restricting the sampling to one plane z = 3.2 Å. Effectively, the unrestricted *M*-sampling of the molecular dynamics trajectories are a posteriori re-centered at z = 3.2 Å. in order to increase the sampling density at the low-temperature equilibrium position (DFT-D locates the G ground microstate at an adsorption height of 3.1 Å). Using this strategy, **Figure S8** shows convergence of the configurational space, monitored through the absolute value of the absolute non-normalized (i.e. without $p_{ref}$) free energy from the PPF method using **Equation 2** and **Equation 3** in the main text after only 40000 MD frames. Finally, we perform symmetry operations in every image shown in **Fig. 5** main text, in order to increase sampling. Importantly, note that these free energies are obtained directly from the 3D PPF matrix instead of a 2D map as explained in **Fig. 4** and **Fig. 5** in the main text.

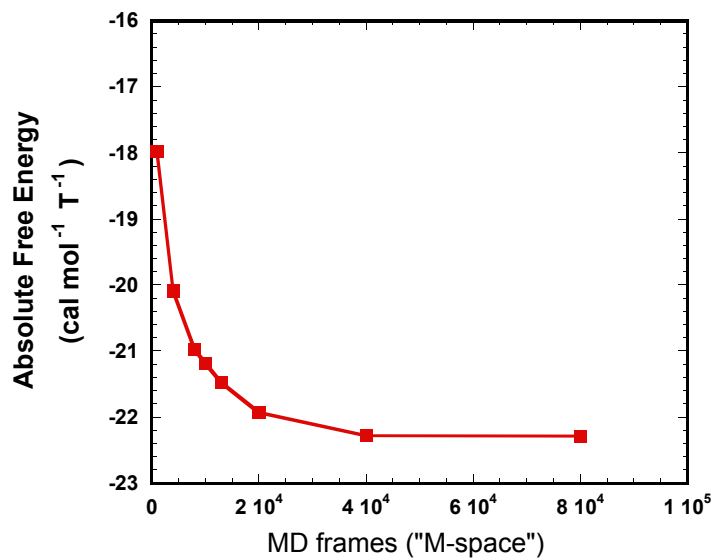

**Figure S9.** Convergence to the full partition function space (*N*-space) as a function of the sampled MD frames (*MD-frames*) at 100 K.

### *ii. Implementation of the sum over states: Effect of exclusion of density of states in sum over energy levels*

For computational efficiency, we use **Equation 3** to sum over energy, rather than configurational space. Effectively, this means that pixels with the same statistical weight are removed, even if they belong to different configurations. This approximation is exact for non-overlapping pixels (see **Fig. S12**) but strongly breaks down at zero Kelvin when overlapping configurations are involved. **Figure S10** illustrates how an image accounting for Rln(12) degeneracy of the G molecular microstate should look like, in comparison with the image created by summing **Equation 3** in energy, rather than in configurational space. As expected, the computational approximation shows an image with the same weight per pixel. We give examples in **Fig. 3** and **Supplementary Note 3**, on how both this computational approximation of the PPF method plays a small role when comparing molecular free energies from the images' free energy (**Equation 2**), provided the area occupied by a molecule can be identified.

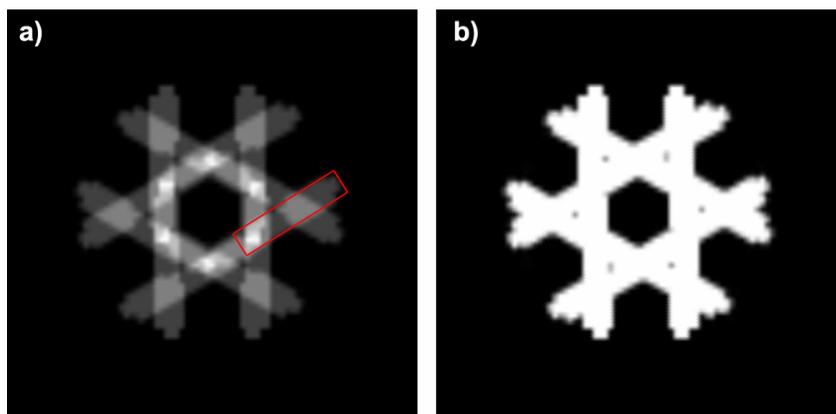

**Figure S10.** Comparison of **a)** an image constructed by averaging 12 degenerate molecular G microstates (a microstate is shown in red) by Equation 1 and one image constructed using 12 degenerate molecular G microstates through in the PPF calculated through the energy level sum implementation used in this work, rather than the exact configurational space Equation 3.

*iii. The interaction energy* **U$_i$**

Finally, the microstate energy in **Equation 3** can be adjusted to include redundant sampling, thus smaller configurational space. First, the microstate energy can be decoupled from fluctuation in the internal degrees of freedom (angles, bond length energies, etc) by exclusively inputting in **U$_i$** the interaction energy of the *ith*-molecular configuration with the pore (so-called INTE energy in CHARMM). In this manner, the G-ground microstate interaction energy is close to -23 kcal mol$^{-1}$. **Figure S11** shows the interaction energy histograms for MD sampling at temperatures of 30 K (*MD-frames*=1x10$^4$), 100 (*MD-frames*=2x10$^4$) and 250K (*MD-frames*=6x10$^4$), revealing that the ground microstate is sampled in all MD simulations, even at temperatures as high as 250 K ( cyan line in Fig. S10). It is also interesting to note that the long high energy Gaussian tail in the 250 K histogram clearly indicates that an open pore system is non-ergodic at 250 K, since molecules have an apparent chance to escape the confinement as described in the main text. For all practical purposes, a pore does constitute a model ergodic system below 145 K.

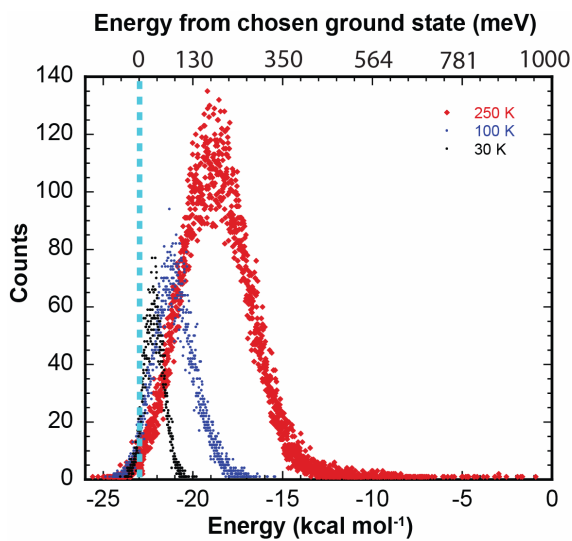

**Figure S11.** The interaction energy histograms at different MD sampling temperatures showing a chosen energetic ground state of the system. Such energies are used in conjunction with Equation 3 in the main text for rendering the PPF method.

### b) *Analytical Partition Function vs. the Projected Partition Function Method.*

A second interest in the PPF method is to ultimately approximate it to analytical partition functions. The major challenge here is to avoid over-counting states close to the numeral precision (1 meV) caused when projecting Boltzmann weights into large pixel sizes. Consider for instance a three-dimensional harmonic oscillator with k=0.004 eV Å$^{-1}$ in a 3D entry matrix of 8 x 8 x 8 Å$^3$ as shown in **Figure S12a**. After MD sampling at 250K with *M*=50000 frames, the potential energy is extracted as shown in **Figure S12b,c**. The time average of the MD itself at 250 K is rendered in **Figure S12d**. The PPF projection method shows that it reproduces much better (more symmetrically) the configurational space than the MD time-average sampling does (**Figure S12e**).

In order to compare the numerical free energies derived from the PPF with the analytical harmonic partition function free energies, **Equation S1**, a reference state must be put forward. In Equation 2 in the main text, the normalized free energy takes as reference state the matrix's maximum value. This corresponds to the center pixel with energy close to zero.

$$z_{vib} = \frac{kT}{v}$$

**Equation S1**. The classical vibration partition function not corrected by Planck's constant.

Consequently, we can put forward a reference state as the translational partition function having as volume the size of the pixel,

$$z_{trans} = V_{pixel}\left(2\pi mkT\right)^{\frac{3}{2}}$$

**Equation S2**. The classical translation partition function not corrected by Planck's constant.

Finally, by using Equation 2 in the main text, we can put forward a normalized analytical free energy as,

$$\overline{F} = -RT \ln\left(\frac{z_{vib}^3}{z_{trans}}\right)$$

**Equation S3**. The normalized expression for the analytical free energy of the harmonic oscillator.

**Figure S12g** shows the comparison between normalized numerical PPF ($p_{ref}$ =max($p$)) from the PPF 3D matrix and analytical free energy expression of **Equation S3**. It shows a fair correlation of both quantities within 2 cal mol$^{-1}$ K$^{-1}$. We assign the observed discrepancies to incomplete configurational space, because of our finite size of the box. We expect future algorithm development for integrating small pixels and large box sizes to provide an excellent correspondence between our numerical PPF method and analytical partition functions.

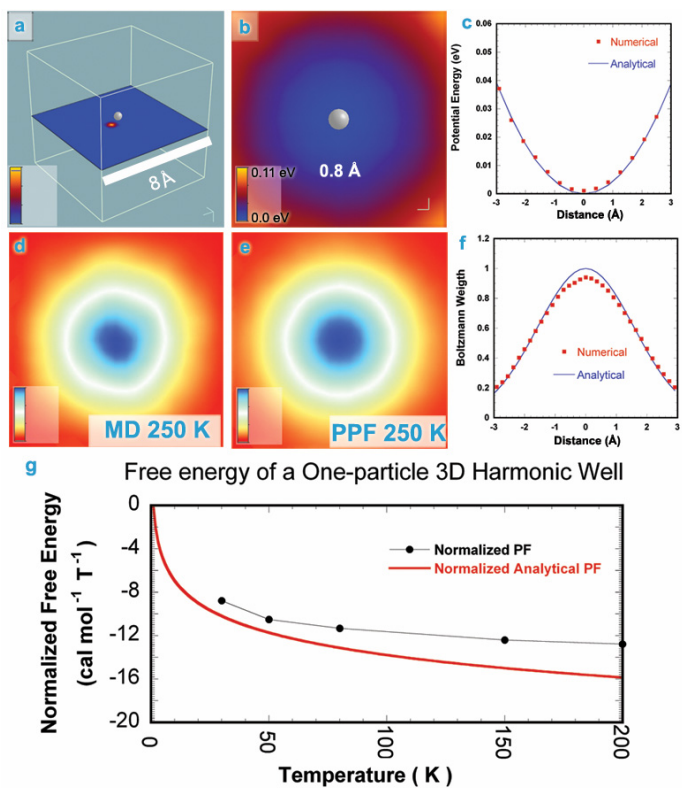

**Figure S12.** Comparison between normalized image free energies between analytical partition functions and numerical PPF for a one-particle 3D harmonic oscillator. a) Box size for the harmonic oscillator. The single atom of mass=1 amu is shown along with a single rendered pixel of size 0.35 Å. b,c) The numerical potential energy of the oscillator. d) MD time-average of the oscillator. e) PPF projection of the oscillator. f) comparison between the analytical Boltzmann weights per pixel and the PPF method. g) The comparison between the image free energies of the PPF projection method and the analytical PF free energies expression in Equation S3.